\documentclass[aps,prl,reprint,groupedaddress,longbibliography]{revtex4-1}
\usepackage{amsmath,amssymb}
\usepackage{graphicx}
\usepackage{braket}
\usepackage{color}

\begin{document}


\title{Anomalous Enhancement of Entanglement Entropy in\\ 
Nonequilibrium Steady States 
 Driven by Zero-Temperature Reservoirs}


\author{Hideaki Hakoshima}
\email{E-mail: hakoshima@as.c.u-tokyo.ac.jp}
\author{Akira Shimizu}
\email{E-mail: shmz@as.c.u-tokyo.ac.jp}
\affiliation{Department of Basic Science, The University of Tokyo, 3-8-1 Komaba, Meguro, Tokyo 153-8902, Japan}
\affiliation{Komaba Institute for Science, The University of Tokyo, 3-8-1 Komaba, Meguro, Tokyo 153-8902, Japan}


\date{\today }

\begin{abstract}
We investigate the size scaling of the entanglement entropy (EE) in  nonequilibrium steady states (NESSs) of a one-dimensional open quantum system with a random potential.  
It models a mesoscopic conductor, composed of a long quantum wire (QWR) with impurities and two electron reservoirs 
at zero temperature.  
The EE at equilibrium obeys the logarithmic law.  
However, in NESSs far from equilibrium the EE grows anomalously fast, obeying the `quasi volume law,' 
although the conductor is driven by the zero-temperature reservoirs.
This anomalous behavior arises from both the far from equilibrium condition and multiple scatterings due to impurities.
\end{abstract}


\maketitle



{\em Introduction.---}
The 
entanglement entropy (EE) 
has been attracting considerable interest
in many fields of physics \cite{amico2008entanglement,eisert2010colloquium,laflorencie2016quantum,jia2008entanglement,laflorencie2006boundary,alet2007valence,kallin2013entanglement,frerot2016entanglement,kitaev2006topological,levin2006detecting,vidal2003entanglement,refael2004entanglement, calabrese2004entanglement,calabrese2009entanglement,bombelli1986quantum,srednicki1993entropy,callan1994geometric,holzhey1994geometric,witten2018aps,
nielsen2002quantum,horodecki2009quantum,
vidal2003entanglement,refael2004entanglement,
kitaev2006topological,levin2006detecting,
ryu2006holographic,page1993information,ryu2006holographic,hotta2015fall,maldacena2016remarks,harlow2016jerusalem,
rajibul2015measuring,kaufman2016quantum,
hastings2007area, pastur2014area,
vidal2003entanglement, refael2004entanglement,calabrese2004entanglement,calabrese2009entanglement,bravyi2012criticality, 
gioev2006entanglement,wolf2006violation,swingle2010entanglement,
movassagh2016supercritical,
movassagh2016supercritical,irani2010ground,
vitagliano2010volume,ramirez2014conformal,salberger2017deformed,d2016quantum,nakagawa2018universality,kim2013ballistic,bardarson2012unbounded,bauer2013area,serbyn2013universal,nandkishore2015many,luitz2015many,serbyn2013local,calabrese2005evolution,bhattacharya2013thermodynamical,
eisler2005entanglement,aschbacher2007non,hoogeveen2015entanglement}, 
including 
quantum information theory,
condensed matter physics 
\cite{jia2008entanglement,laflorencie2006boundary,alet2007valence,kallin2013entanglement,frerot2016entanglement,kitaev2006topological,levin2006detecting,
vidal2003entanglement,refael2004entanglement},  
quantum field theories \cite{bombelli1986quantum,srednicki1993entropy,callan1994geometric,holzhey1994geometric,witten2018aps,calabrese2004entanglement,calabrese2009entanglement},
and quantum gravity \cite{page1993information,ryu2006holographic,hotta2015fall,maldacena2016remarks,harlow2016jerusalem}.
This is because
the EE is found useful 
not only for 
quantifying the resources for quantum information tasks \cite{nielsen2002quantum,horodecki2009quantum}
but also for analyzing 
physical properties such as the central charge \cite{vidal2003entanglement,refael2004entanglement}, 
topological order \cite{kitaev2006topological,levin2006detecting}, 
many-body localization \cite{serbyn2013universal,nandkishore2015many,luitz2015many,serbyn2013local},
and the Bekenstein-Hawking entropy \cite{page1993information,ryu2006holographic,hotta2015fall,maldacena2016remarks,harlow2016jerusalem}.
Recent experiments have succeeded in measuring the EE \cite{rajibul2015measuring,kaufman2016quantum}.
In particular, 
intensive studies have been conducted on the asymptotic size dependence 
of the EE $S_L$ of the ground states of one-dimensional systems.
In this case, $S_L$
quantifies the 
entanglement between a subsystem of length $L$ and the rest of the system. 
In many systems with natural Hamiltonians, 
the {\em area law} $S_L=O(1)$ \cite{hastings2007area, pastur2014area}
and the {\em logarithmic law} $S_L=O(\ln{L})$
\cite{vidal2003entanglement, refael2004entanglement,calabrese2004entanglement,calabrese2009entanglement,bravyi2012criticality, 
gioev2006entanglement,wolf2006violation,swingle2010entanglement}
were found,
%
in consistency with thermodynamics (i.e., $S_L=o(L)$ at zero temperature).
Larger $S_L$ was found only in artificial 
 toy models \cite{movassagh2016supercritical,irani2010ground,vitagliano2010volume,ramirez2014conformal,salberger2017deformed}.

These numerous works have studied the EE of 
the ground states or other energy eigenstates, 
almost all of which are equilibrium states according to 
the eigenstate-thermalization hypothesis \cite{jensen1985statistical,deutsch1991quantum,srednicki1994chaos,rigol2008thermalization,kim2014testing,d2016quantum,tasaki2016typicality,iyoda2017fluctuation}.
A natural question is
how the EE behaves in nonequilibrium states.
With regard to the thermalization processes in isolated systems,
the time evolution of the EE was studied 
in terms of condensed matter \cite{d2016quantum,nakagawa2018universality,kim2013ballistic,bardarson2012unbounded,bauer2013area,serbyn2013universal,nandkishore2015many,luitz2015many,serbyn2013local}, 
quantum field theories and quantum gravity \cite{calabrese2005evolution,bhattacharya2013thermodynamical}.
In these systems, 
$S_L \leq O(S_L^{\rm eq}(E))$ throughout the evolution,
where $S_L^{\rm eq}(E)$ is the equilibrium entropy at energy $E$
of the system.
Regarding NESSs, 
which are fundamental states in nonequilibrium physics
\cite{Lax1960,SandT1971,zubarev1974nonequilibrium,jou1996extended,OandP1998,SandY2010,AS2010,sagawa2011geometrical,lieb2013entropy,sasa2014possible,tasaki2001nonequilibrium},
their EE was studied for certain systems
\cite{eisler2005entanglement,aschbacher2007non,hoogeveen2015entanglement}.
As in the case of thermalization processes, 
it was shown that 
$S_L = O(S_L^{\rm eq}(E))$ or 
$S_L = O(\sum_\nu S_L^{\rm eq}(T^\nu_{\rm res}))$, 
where $T^\nu_{\rm res}$ is the temperature of the $\nu$th reservoir.
However, 
 this is because 
 the systems of these NESSs are invariant under spatial translation,
and consequently the NESSs are basically the boosts of equilibrium states.
In contrast, the NESSs observed in common experiments are
those of systems with symmetry-breaking scatterings, e.g., 
by impurities, rough walls, or phonons,
which define a particular rest frame.
Multiple scatterings by such scatterers make NESSs nontrivial, i.e., 
much different from the boosts of equilibrium states.

\begin{figure}
\centering
\includegraphics[width=0.46\textwidth]{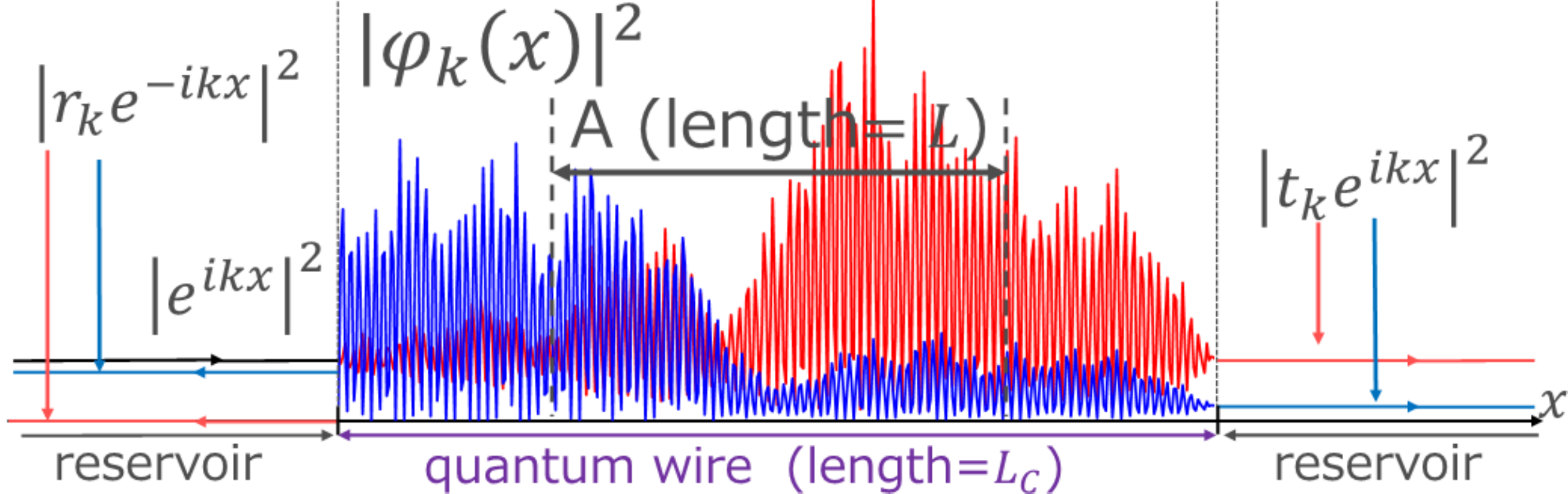}
\caption{
(Color online) Two examples of $\varphi_k(x)$ for $W=0.08$,
with $k=0.82128\cdots$ (red) on a resonant point,
and $k=0.82479\cdots$ (blue) at the middle point
between this resonant and adjacent off-resonant points.
$|\varphi_k(x)|^2$ is plotted in the QWR ($|x| \leq L_{\rm C}/2$)
and the square of the incoming and outgoing waves are plotted 
separately in the reservoirs ($|x| > L_{\rm C}/2$).
}
\label{fig:wf}
\end{figure}

In this letter, we study $S_L$ 
of NESSs in a one-dimensional 
mesoscopic conductor 
with impurities 
\cite{sakaki1989quantum,imry2002introduction,tilke2003quantum,datta1997electronic,shimizu1998interacting,SK2000,blanter2000shot,tasaki2001nonequilibrium},
which is a long 
quantum wire (QWR) 
connected to 
two electron reservoirs 
of zero temperature ($T^\nu_{\rm res} = 0$). 
The difference 
$\Delta \mu := \mu^+-\mu^-$ 
in the chemical potentials 
$\mu^\pm$ of
the reservoirs induces a steady current $J$ in the QWR, 
and a NESS is realized.
While $S_L^{\rm eq}=O(\ln L)$ at equilibrium, 
we find that, in nontrivial NESSs far from equilibrium 
(as defined by (\ref{eq:ffeq}) below), 
\begin{equation}
S_L
=
\eta(L) L |\Delta k_F| + O(\ln L)
\
\mbox{ for } 1 \ll L \leq L_{\rm C}.
\label{eq:W>0,Dm>0}
\end{equation}
Here, 
$\Delta k_F$ is the difference in the Fermi wavenumbers of the reservoirs,
$L_{\rm C}$ is the length of the QWR,
and $\eta(L)$ is a function of $L$ with the following properties:
When $1 \ll L \leq L_{\rm C}$, 
(i) $\eta(L)$ is independent of $\Delta k_F$,
(ii) gradually decreases with increase in $L$, 
(iii) and
\begin{equation}
a \le \eta(L) \lesssim 2a
\
\mbox{ for } 1 \ll L \leq L_{\rm C},
\label{eq:a<eta<2a}
\end{equation}
where $a$ is a positive constant independent of $L$ or $\Delta k_F$.
Since 
$
S_L
\geq 
a L |\Delta k_F| + O(\ln L)
$,
we call Eq.~(\ref{eq:W>0,Dm>0}) 
the {\em quasi volume law}.
Consequently, 
$S_L > O(\sum_\nu S_L^{\rm eq})$
in contrast to $S_L \leq O(\sum_\nu S_L^{\rm eq})$
in the previous cases 
\cite{d2016quantum,nakagawa2018universality,kim2013ballistic,bardarson2012unbounded,bauer2013area,serbyn2013universal,nandkishore2015many,luitz2015many,serbyn2013local,calabrese2005evolution,bhattacharya2013thermodynamical,eisler2005entanglement,aschbacher2007non,hoogeveen2015entanglement}.
{\em Both} the far from equilibrium condition
and multiple scatterings that break the translational symmetry 
are necessary for this anomalous enhancement of $S_L$.
{\em Setup.---}
We consider a long QWR (conductor)
\cite{sakaki1989quantum,imry2002introduction,tilke2003quantum,datta1997electronic,shimizu1998interacting,SK2000,blanter2000shot,tasaki2001nonequilibrium} 
connected to two
electron reservoirs of 
zero temperature.
Although real reservoirs are usually two-dimensional, 
the total system
can be mapped to a one-dimensional system 
\cite{shimizu1998interacting,SK2000}.
If many-body interactions are negligible, 
its effective Hamiltonian is given by
\begin{align}
 \hat{H}_{\rm tot}
:=
-\sum_x (\hat{c}^{\dagger}_x \hat{c}_{x+1}+h.c.)
+\sum_{ |x| \leq L_{\rm C}/2}v_{x}\hat{c}^{\dagger}_{x}\hat{c}_{x}.
\label{eq:H}
\end{align}
Here, $\hat{c}^{\dagger}_x$ and $\hat{c}_x$ are the creation and annihilation operators of an electron at site $x$ ($\in \mathbb{Z}$), respectively.
 In the QWR 
of length $L_{\rm C}$ centered at $x=0$ (see Fig.~\ref{fig:wf} and  Fig.~S1 of \cite{SM} for details),
a Gaussian random potential $v_x$ of impurities exists
(with a vanishing average \cite{SM}).
Its strength is characterized by 
the standard deviation $W$ of $v_x$.

We require that a single-particle state
should be either a 
scattering state $\varphi_k(x)$ (Fig.~\ref{fig:wf})
with incoming wavenumber $k$ ($-\pi < k \leq \pi$) and 
energy $\varepsilon_k=-2\cos{k}$,
or a bound state $\varphi_b(x)$ 
with a quantum number $b$ 
and energy $\varepsilon_b$ ($|\varepsilon_b| > 2$).
According to
the standard model of mesoscopic conductors
\cite{datta1997electronic,imry2002introduction,tilke2003quantum,lesovik1989quantum,buttiker1990scattering,li1990low,shimizu1992effects,shimizu1993effects,blanter2000shot,landauer1957spatial,landauer1987electrical,buttiker1985generalized,
shimizu1998interacting,SK2000,SM},
the quantum state at zero temperature 
of the total system  is 
a pure quantum state  $|\Psi_{\rm tot}\rangle$ such that 
$\varphi_k(x)$ with $-k_F^-\le k \le k_F^+$
and 
$\varphi_b(x)$ with $\varepsilon_b < -2$ 
are occupied by electrons \cite{SM}. 
Here, $k_F^+$ ($k_F^-$) is the Fermi wavenumber 
of the left (right) reservoir, i.e.\ 
$
\varepsilon_{k_{F}^{\pm}}
=\mu^{\pm}
= \overline{\mu} \pm \Delta \mu/2,$
where $\overline{\mu} := (\mu^++\mu^-)/2$.
Without loss of generality, we assume that $\Delta \mu  \geq 0$,
and hence $\Delta k_F := k_F^+ - k_F^- \geq 0$.
A NESS is realized when $\Delta \mu  > 0$.

{\em Entropy.---}
Assuming $|\Psi_{\rm tot}\rangle$,
we explore its EE.
We take 
a subsystem A 
$:= [-L/2, L/2]$
of length $L$ at the center of the QWR (see Fig.~\ref{fig:wf} and  Fig.~S1 of \cite{SM} for details), 
and consider 
the von Neumann entropy 
$S_L :=-\mathrm{Tr}[\hat{\rho}_L \ln \hat{\rho}_L]$ 
of the reduced density operator $\hat{\rho}_L$ of A.
 It is known that $S_L$ at equilibrium 
agrees with the thermodynamics entropy \footnote{
This point is a direct consequence of the following facts:
(a) If the total system is in an equilibrium state then its subsystem, which is much smaller than the total system, is in the (grand)canonical Gibbs state.
(b) The von Neumann entropy of the (grand)canonical Gibbs state agrees with its thermodynamic entropy; see, e.g., L.D Landau, E.M Lifshitz {\it Statistical Physics, Part I} (Pergamon Press, 1980).
}.
Since $|\Psi_{\rm tot}\rangle$ is a pure quantum state, 
$S_L$ is also the EE 
that quantifies the 
entanglement between A 
and the rest of the system,
either at equilibrium or nonequilibrium.

For each $|\Psi_{\rm tot}\rangle$, which is determined by 
$k_F^\pm$ and the impurities,
we examine the $L$ dependence of $S_L$. 
We are most interested in $S_L$ 
{\em in the QWR}
(i.e.\ $S_L$ for $L \leq L_{\rm C}$),
in which the quantum state in a NESS differs significantly from that in 
an equilibrium state.


{\em Nontrivial NESSs.---}
From the electron-hole symmetry, 
we can limit ourselves, without loss of generality,
 to the lower half of the band, 
$-2 \leq \varepsilon_k \leq 0$.
Furthermore, since we are not interested 
in any specific effect of the band edge $\varepsilon_k =-2$
or the band center $\varepsilon_k =0$,  
we take $-1.7 \leq \mu^\pm \leq -0.8$.

We exclude the case of a short QWR, 
$L_{\rm C} \sim 1$,
because such a QWR is actually a quantum dot, 
for which we cannot discuss the $L$ dependence of $S_L$ for $L\le L_{\rm C}$.
We therefore study the case of $L_{\rm C} \gg 1$.
We take $L_C=401$ in the numerical calculations \cite{SM}.

Since we assume zero temperature, 
the dimensionless conductance 
\cite{imry2002introduction}
$G := (J/\Delta \mu)/(e^2/2 \pi \hbar)$ 
(which is a nonlinear one; see below) 
is simply the average value
$G = {1 \over \Delta \mu} \int_{\mu^-}^{\mu^+} |t_k|^2 d\varepsilon_k$
of the transmittance $|t_k|^2$
in $\mu^- \leq \varepsilon_k \leq \mu^+$.
Obviously, $0 \leq G \leq 1$. 
%
Since $L_{\rm C}$ is finite and impurities are absent in the reservoirs, 
the Anderson localization 
\cite{anderson,abrahams1979scaling,anderson1980new,lee1985disordered}
is incomplete, 
i.e., 
the localization length $\xi$
(defined for the hypothetical case $L_{\rm C} = \infty$)
can exceed $L_{\rm C}$.

As $W$ is increased, $\xi$ and $G$ decrease on the whole 
\cite{anderson,abrahams1979scaling,anderson1980new,lee1985disordered}.
%
%
When $L_{\rm C} \gg \xi$, 
the system would be almost an insulator, and $G \simeq 0$. 
Hence, almost no current would flow even when finite $\Delta \mu$ is applied.  
On the other hand, when $L_{\rm C}$ is much shorter than the mean free path $(<\xi)$, 
the electrons would not suffer scatterings, and $G \simeq 1$. 
The NESS in this case would be almost a boost of an equilibrium state (even in the presence of interactions between electrons \cite{shimizu1996landauer}). 
In these cases, it is obvious that 
$S_L$ scales as in equilibrium.
We  therefore focus on the intermediate regime where $L_{\rm C}$ is comparable to 
$\xi$, for which $G$ takes an intermediate value,
such as $0.3 \lesssim G \lesssim 0.7$.
We call the NESSs in this case {\em nontrivial}, 
and take $W$ so as to satisfy this condition.

For such NESSs,
multiple scatterings by impurities are crucial,
and the wavefunctions $\varphi_k(x)$ 
have complicated shapes as shown in Fig.~\ref{fig:wf}.
Since $k$ takes a continuous value, there are infinitely many states, 
including those nearly localized in the QWR (such as the red one)
and those penetrating into the QWR (such as the blue one).
%
%
Consequently, 
$|t_k|^2$
varies rapidly as a function of $k$, as shown in Fig.~S2 of \cite{SM}, 
where each peak 
indicates the resonant tunneling through a nearly-localized state.

{\em Far from equilibrium.---}
Since $-\pi < k \leq \pi$ and 
the number of resonant states $\simeq L_{\rm C}$, 
the average distance $\Delta k_{\rm peak}$ between the peaks of $|t_k|^2$
is roughly 
$\Delta k_{\rm peak} \simeq 2 \pi/L_{\rm C}$.
When $\Delta k_F \ll \Delta k_{\rm peak}$,
the current-voltage characteristic is linear, i.e.\ 
$G$ is independent of $\Delta \mu$ 
(while $G$ depends on $\overline{\mu}$).
In this regime, the NESS is close to equilibrium.
When $\Delta k_F$ is increased to $\Delta k_F \sim \Delta k_{\rm peak}$,
$G$ depends sensitively on $\Delta \mu$ (and $\overline{\mu}$), 
as shown in Fig.~S3 of \cite{SM},
because only a small number of peaks in 
$k_F^- \leq k \leq k_F^+$ contribute to the conduction,
which reflects the characteristics of the individual peaks.
When $\Delta k_F$ is further increased to 
$\Delta k_F \gg \Delta k_{\rm peak}$,
the dependence on $\Delta \mu$ (and $\overline{\mu}$) becomes weak
because many peaks and dips of $|t_k|^2$ contribute to 
the conduction.
When $\Delta k_F = O(1)$, such a regime is always achieved 
for sufficiently large $L_{\rm C}$ because 
$\Delta k_{\rm peak} \simeq 2 \pi/L_{\rm C}$.
We call this regime in which 
\begin{equation}
\Delta k_F = O(1) \gg \Delta k_{\rm peak}
\label{eq:ffeq}
\end{equation}
{\em far from equilibrium}.
We will show the quasi volume law (\ref{eq:W>0,Dm>0})
 for nontrivial NESSs far from equilibrium.

{\em Relation to number and current.---}
Let 
$\delta N_L^2 :=\braket{(\hat{N}_L - \braket{\hat{N}_L})^2}$,
which is the fluctuation of the particle number 
$\hat{N}_L :=\sum_{|x| \leq L/2}\hat{c}^{\dagger}_x \hat{c}_{x}$ in A,
where 
$
\braket{\bullet}
:=
\braket{\Psi_{\rm tot} | \bullet | \Psi_{\rm tot}}
$.
In \cite{SM}, we prove the inequality
\begin{align}
\delta N_L^2 \le S_L\le 1+c(\ln L)\delta N_L^2,
\label{eq:dNinequality}
\end{align}
where $c$ is a positive constant independent of $L$.
From this inequality, 
it is sufficient to show the quasi-volume law for $\delta N_L^2$
instead of $S_L$.
We will analyze $\delta N_L^2$
to see the mechanism and the order of magnitude, 
and calculate $S_L$ numerically to find the 
magnitude, of the anomalous enhancement.

To calculate $\delta N_L^2$, 
we neglect small contributions from the bound states, and 
use the identity 
\cite{SM}:
\begin{align}
\delta N_L^2 
= 
\iint_{\Omega_{\Delta\mu}} dk_1dk_2 R_L^W(k_1,k_2).
\label{eq:dN=intF}
\end{align}
Here, 
the region of the integral $\Omega_{\Delta\mu}$
is such that $k_1$ is occupied by an electron while $k_2$ is empty.
Furthermore, 
\begin{equation}
R_L^W(k_1,k_2)
:=
|\Delta J_L^{p q}|^2 
\big/ 
[16 \sin^2 \left(p/2 \right) \sin^2  \left( q/2 \right)],
%
%
\label{def:GL}
\end{equation}
where
$p:=k_1+k_2$ and $q := k_1-k_2$.
Here,
$
\Delta J_L^{p q} 
:= J_{k_1k_2}(L/2) - J_{k_1k_2}(-L/2)
$,
where 
$J_{k_1k_2}(x+1/2)$ is the $k$ representation of the current 
on the bond at $x+1/2$; 
$ 
J_{k_1k_2}(x+1/2) 
=
i 
[\varphi^*_{k_1}(x+1)\varphi_{k_2}(x) - \varphi^*_{k_1}(x)\varphi_{k_2}(x+1)].
$ 
When $p=0$ or $q=0$, 
where the denominator of $R_L^W$ vanishes,
the numerator 
$|\Delta J_L^{p q}|^2$
also vanishes
\cite{SM}.
Consequently, $R_L^W$ is finite everywhere in $\Omega_{\Delta\mu}$.
Identity (\ref{eq:dN=intF}) means that 
$\delta N_L^2$ in A is determined by the net current flowing into A
through its edges, 
$x = \pm L/2$.

In the following discussion, identity (\ref{eq:dN=intF}) plays a crucial role.
It decomposes the parameter dependence into two, 
$\Omega_{\Delta\mu}$ and $R_L^W$, 
which depend on $(\Delta\mu, \overline{\mu})$ and 
$(W,L,L_{\rm C})$, respectively.
We will show that $R_L^W$ takes large values of $O(L^2)$ 
in certain regions in the $k_1$-$k_2$ plane,
whereas $\Omega_{\Delta\mu}$
determines which parts are extracted from such regions.

{\em Behavior of $R_L^W$ (summarized in Table \ref{tabel:R}).---}
For $q=O(1/L)$, 
the denominator of Eq.~(\ref{def:GL}) becomes as small as $O(1/L^2)$,
and consequently
the typical value of $R_L^W$ becomes as large as $O(L^2)$.
In contrast, for $p=O(1/L)$, 
$R_L^W=O(1)$ when $W=0$, because then the numerator 
also becomes $O(1/L^2)$.

When $W>0$, however, 
the difference 
between the forward-scattering part (small $|q|$) and the backward-scattering part  (small $|p|$) is obscured
because the impurities scatter the electrons back and forth.
As a result, $R_L^W$ becomes $O(L^2)$ not only at $q=O(1/L)$ but 
also at $p=O(1/L)$, as plotted in Figs.~S5 and S6 of \cite{SM}.


Since $R_L^W=O(1)$ in the other regions in $\Omega_{\Delta\mu}$,
the $L$ dependence of $\delta N_L^2$ is determined by the integral around $q\simeq 0$ and $p\simeq 0$.
We therefore focus on the regions $|q| \leq \epsilon$ and 
$|p| \leq \epsilon$, where $\epsilon$ is a positive constant of $O(1)$.

\begin{table}
\begin{minipage}{0.49\hsize}
\centering
\caption{Typical value of\\ $R_L^W$ (and $\tilde{R}_L^W$)}
\label{tabel:R}
\begin{tabular}{|l|c|c|} \hline
& $W=0$ & $W>0$\\ \hline
$q=O(1/L)$ &$O(L^2)$&$O(L^2)$ \\ \hline
$p=O(1/L)$ &$O(1)$&$O(L^2)$ \\ \hline
\end{tabular}
\end{minipage}
\begin{minipage}{0.49\hsize}
\centering
\caption{Areas between $q$ and $q + dq$, 
and $p$ and $p + dp$.} 
\label{tabel:Omega}
\begin{tabular}{|l|c|c|} \hline
& $\Delta\mu=0$ & $\Delta\mu>0$\\ \hline
$q\simeq0$ &$|q| dq$&$|q| dq$ \\ \hline
$p\simeq0$ &$|p| dp$&$|p-\Delta k_F| dp$ \\ \hline
\end{tabular}
\end{minipage}
\centering
\caption{$L$ dependence of $S_L$ and $\delta N_L^2$ ($L\le L_C$)}
\label{tabel:SL}
\begin{tabular}{|c|l|l|} \hline
& $W=0$ & $W>0 $ \\ \hline
$\Delta\mu=0$ &[A] $O(\ln{L})$&[C] $O(\ln{L})$ \\ \hline
$\Delta\mu>0$ &[B] $O(\ln{L})$&[D] 
\begin{tabular}{l}quasi volume law  (for nontrivial \\ NESSs far from equilibrium) 
\end{tabular} \\ \hline
\end{tabular}
\end{table}

\begin{figure}[htp]
\centering
\includegraphics[width=0.46\textwidth]{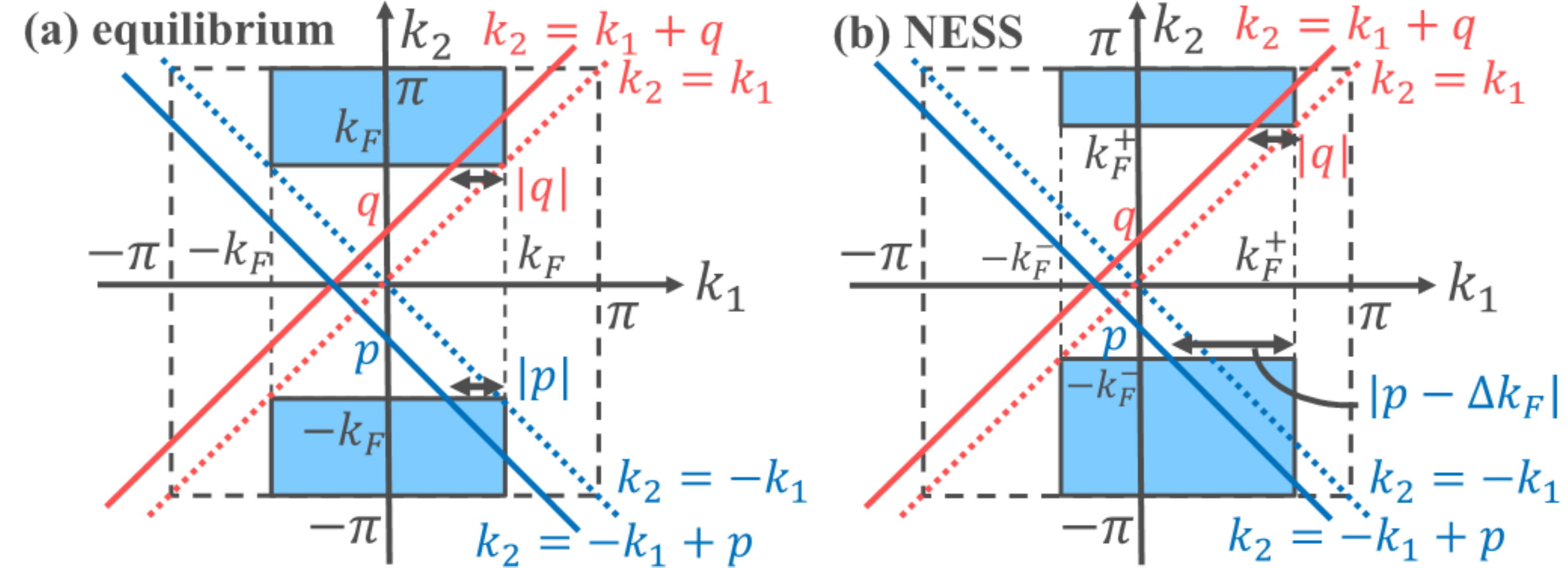}
\caption{(Color online) The region $\Omega_{\Delta\mu}$ of the integral for
(a) an equilibrium state ($\Delta \mu = 0$),
and (b) a NESS ($\Delta \mu > 0$).
}
\label{fig:region}
\end{figure}

{\em Behavior of $\Omega_{\Delta\mu}$ (Table \ref{tabel:Omega}).---}
For an equilibrium state, 
the region of the integral $\Omega_{\Delta\mu}$ is shown 
in Fig.~\ref{fig:region}(a).
As discussed above, we focus on the regions $q \simeq 0$ and 
$p \simeq 0$.
Then, the area of the portion of $\Omega_{\Delta\mu}$
between $q$ and $q + dq$ ($p$ and $p + dp$) is $|q| dq$ ($|p| dp$).

For a NESS, 
$\Omega_{\Delta\mu}$ shifts toward the direction of $+45^\circ$, 
as shown in Fig.~\ref{fig:region}(b).
As a result, the area between $p$ and $p + dp$ 
becomes $|p- \Delta k_F| dp$,
while that between $q$ and $q + dq$ 
remains the same as in the equilibrium case.
This means that in nonequilibrium,  more contributions are extracted from 
$R_L^W$ of small $|p|$.



{\em Clean case (Table \ref{tabel:SL} [A] and [B]).---}
When impurities are absent ($W=0$), 
we evaluate Eq.~(\ref{eq:dN=intF}) 
using Tables \ref{tabel:R} and \ref{tabel:Omega},
as 
$\delta N_L^2=\int_{-\epsilon}^{\epsilon}  R_L^W |q|dq +O(1)$
for either equilibrium ($\Delta \mu =0$) or nonequilibrium ($\Delta \mu >0$).
A simple power-counting argument 
estimates the integral as 
$O(1/L)$ [interval where $q = O(1/L)$]
$\times$ 
$O(1/L)$ [from $|q|$]
$\times$ 
$O(L^2)$ [from $R_L^W$, Table \ref{tabel:R}]
$= O(L^0)$.
Actually, the rigorous argument in Sec.IV of \cite{SM} gives $\delta N_L^2 = O(\ln L)$.
From inequality (\ref{eq:dNinequality}), this indicates the logarithmic law $S_L=O(\ln L)$, 
which is confirmed numerically (Fig.~S4 of \cite{SM}).
This agrees with the previous results for $\Delta \mu =0$
\cite{gioev2006entanglement,wolf2006violation}.
The same result holds for $\Delta \mu >0$ (case [B])
because a NESS for $W=0$ is basically the boost of an equilibrium state \cite{SM}.


{\em Equilibrium with random potential (Table \ref{tabel:SL} [C]).---}
When the random potential is introduced ($W>0$), 
$R_L^W=O(L^2)$ not only at $q \simeq 0$ but also at $p \simeq 0$
 (Table \ref{tabel:R}).
Hence, unlike the clean case, 
both the regions contribute to the $L$ dependence of $\delta N_L^2$.
Then, for $\Delta \mu =0$, 
we evaluate Eq.~(\ref{eq:dN=intF}) 
using Table \ref{tabel:Omega},
as
\begin{equation}
\delta N_L^2
=\int_{-\epsilon}^{\epsilon}  \tilde{R}_L^W |q|dq 
+\int_{-\epsilon}^{\epsilon}  \tilde{R}_L^W|p| dp +O(1).
\label{eq:dN_eq}
\end{equation}
Here, $\tilde{R}_L^W$ is 
the average of $R_L^W$ over $k_1$ around $q,p\simeq 0$,
which has the same typical value as $R_L^W$ (Table \ref{tabel:R}).
Since both the integrals 
give $O(\ln L)$ for the same reason as that of the clean case,
we again have $\delta N_L^2 =O(\ln L)$ and $S_L=O(\ln L)$
(see \cite{SM} for details). 
This is demonstrated in Fig.~\ref{fig:W>0,Dm>0} (red symbols) and Fig.~S4 of \cite{SM}.
\begin{figure}
\centering
\includegraphics[width=0.46\textwidth]{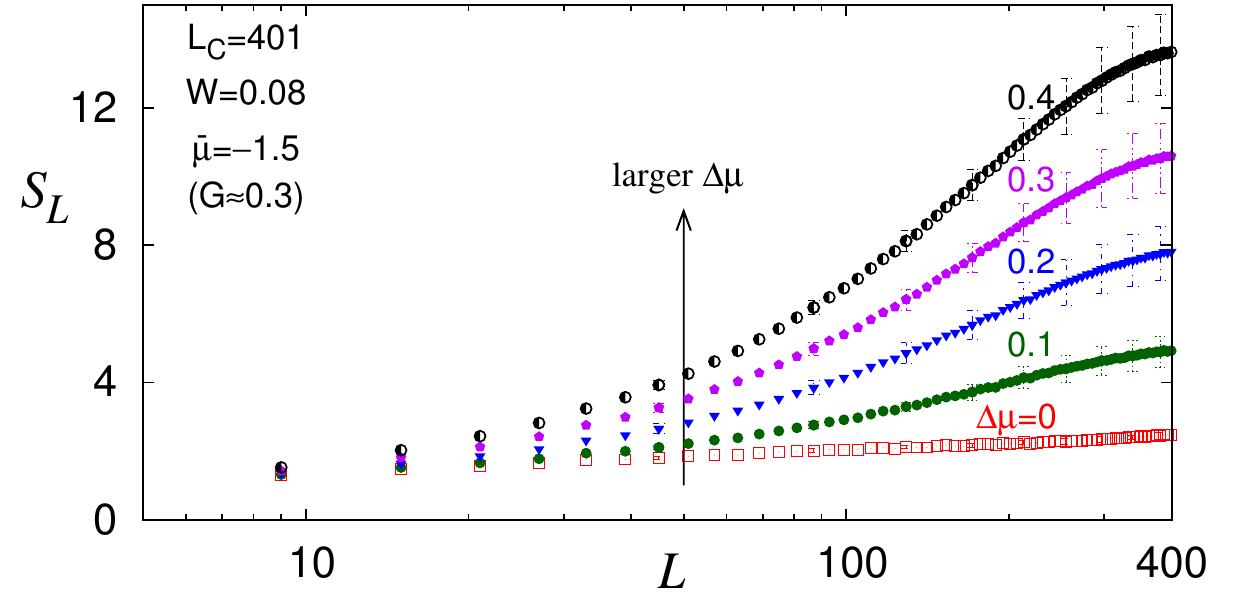}
\caption{
(Color online) The average and standard deviation (error bars), over $10$ random samples,
of $S_L$ 
against $\log L$ 
for $L \leq L_{\rm C}$
and for various values of $\Delta \mu$.
}
\label{fig:W>0,Dm>0}
\end{figure}


{\em Nontrivial NESSs (Table \ref{tabel:SL} [D]).---}
Figure \ref{fig:W>0,Dm>0} also shows that,
with increasing $\Delta \mu$, 
$S_L$ grows considerably, and it cannot be fitted by a
linear function of $\ln L$.
This happens by the following mechanism.

When $\Delta \mu >0$, 
the area between $p$ and $p + dp$ becomes $|p- \Delta k_F| dp$
(Table \ref{tabel:Omega}).
Hence, 
Eq,~(\ref{eq:dN_eq}) changes to 
\begin{equation}
\delta N_L^2
=
\int_{-\epsilon}^{\epsilon}  \tilde{R}_L^W |q|dq +\int_{-\epsilon}^{\epsilon}  \tilde{R}_L^W |p-\Delta k_F|dp +O(1).
\end{equation}
%
The first integral gives $O(\ln L)$ for the same reason as those of cases [A]-[C].
For the second integral, 
we can rewrite it, 
by taking $\epsilon = O(1) < \Delta k_F$, as 
$\Delta k_F \int_{-\epsilon}^{\epsilon} \tilde{R}_L^W dp+\int_{-\epsilon}^{0} \tilde{R}_L^W |p|dp-\int_{0}^{\epsilon}  \tilde{R}_L^W |p|dp.$ 
The last two terms give $O(\ln L)$ as before (though expected to cancel each other out after the random average).
In contrast, the integral $\int_{-\epsilon}^{\epsilon} \tilde{R}_L^W dp$ basically gives $O(L)$, because 
$O(1/L)$ [interval where $p = O(1/L)$]
$\times$ 
$O(L^2)$ [from $\tilde{R}_L^W$, Table \ref{tabel:R}]
$= O(L)$.

This indicates that $S_L= O(L)$. However, 
there is a slight correction 
because the backward scatterings become weaker with increasing $L$, i.e.\ as the edges 
($x=\pm L/2$) of A approach the edges ($x=\pm L_{\rm C}/2$) of the QWR.
Consequently, the typical value of $R_L^W$ grows slightly slower than $O(L^2)$ with increasing $L$, as shown in Figs.~S5 and S6 of \cite{SM}.
By representing this effect with a gradually decreasing function $\eta(L)$,
we arrive at Eq.~(\ref{eq:W>0,Dm>0}).

\begin{figure}
\centering
\includegraphics[width=0.46\textwidth]{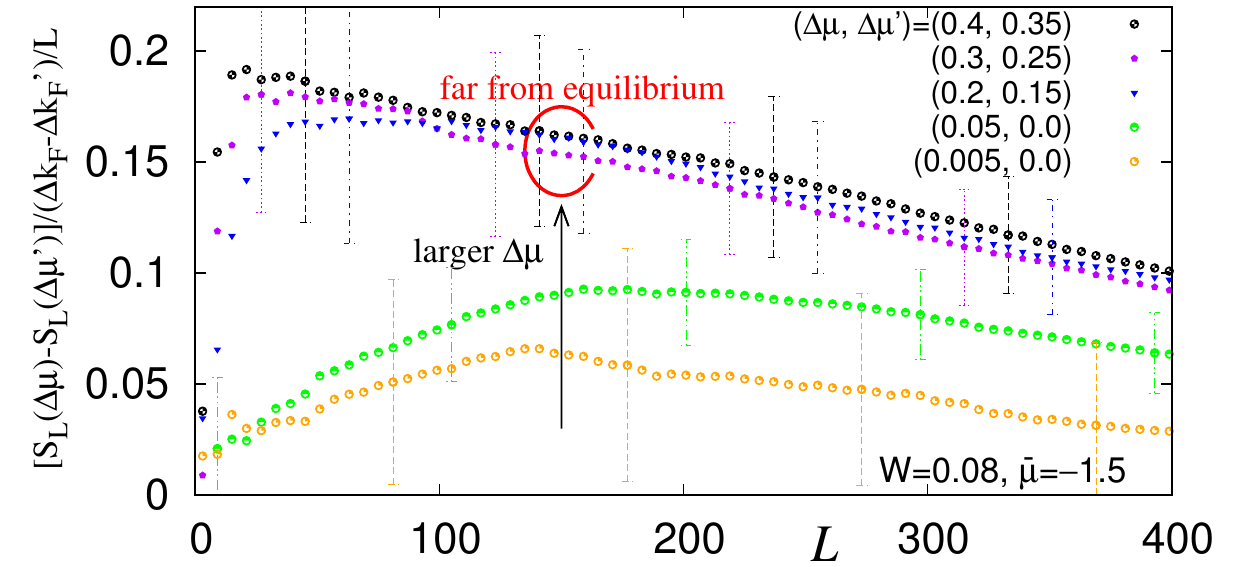}
\caption{
(Color online) The average and standard deviation (error bars), over $10$ random samples,
of the r.h.s.\ of Eq.~(\ref{eq:sabun})
for various values of $(\Delta \mu, \Delta \mu')$,
for the same values of $W$ and $\overline{\mu}$ as in Fig.~\ref{fig:W>0,Dm>0}.
}
\label{fig:eta}
\end{figure}

To confirm the quasi volume law, we investigate whether $\eta(L)$ 
has properties (i)-(iii), mentioned earlier 
around inequality (\ref{eq:a<eta<2a}), 
for nontrivial NESSs far from equilibrium.
We note that the $O(\ln L)$ term in 
the r.h.s.\ of Eq.~(\ref{eq:W>0,Dm>0})
should be relatively insensitive to $\Delta k_F$.
Hence, 
by subtracting Eq.~(\ref{eq:W>0,Dm>0}) at $\Delta \mu$
from that at $\Delta \mu'$, 
where $|\Delta \mu - \Delta \mu'| \ll 1$, 
we expect
%
%
\begin{equation}
\eta(L)
\simeq
[S_L(\Delta \mu) - S_L(\Delta \mu')]/(\Delta k_F - \Delta k'_F)L,
\label{eq:sabun}
\end{equation}
for $1 \ll L \leq L_{\rm C}$.
We plot the r.h.s.\ of Eq,~(\ref{eq:sabun})
for various values of $(\Delta \mu, \Delta \mu')$ in Fig.~\ref{fig:eta}.
When $(\Delta \mu, \Delta \mu')$ is small 
(close to equilibrium),
none of the properties (i)-(iii)
is satisfied \footnote{
When $\Delta k_F \lesssim \Delta k_{\rm peak}$,
the integral 
$
\int \tilde{R}_L^W dp \propto (\Delta k_F)^2
$
as $\Delta k_F \to 0$, 
and hence the r.h.s.\ of Eq.~(\ref{eq:sabun}) 
should decrease with decreasing $\Delta k_F$,
as seen from Fig.~\ref{fig:eta}.
}.
However, when $(\Delta \mu, \Delta \mu')$ is large, 
all the properties 
are satisfied.
For example, property (iii) is satisfied with 
$a \simeq 0.1$ \cite{SM}.
We have thus confirmed the quasi volume law
for nontrivial NESSs far from equilibrium.


These results are summarized in Table \ref{tabel:SL}.
It clearly shows that 
{\em the quasi volume law is peculiar to nontrivial NESSs \cite{invitedtalk}.}

{\em Scaling in reservoirs.---}
When $L$ is increased to $L > L_{\rm C}$, 
the $O(L^2)$ components of $R_L^W$ 
at small $|p|$ cease to grow with increasing $L$,
as shown in Figs.~S8 and S9 of \cite{SM},
because impurities are absent in the reservoir regions,
whereas those at small $|q|$ continue to grow.
%
Consequently, the logarithmic law is recovered with an offset value: 
\begin{equation}
S_L
=
 \eta(L_{\rm C})L_{\rm C} |\Delta k_F| + O(\ln L) 
\ \mbox{ for $L > L_{\rm C}$},
\label{eq:res.W>0,Dm>0}
\end{equation}
as demonstrated in Fig.~S10 of \cite{SM}.

In summary, 
at equilibrium, $S_L$ 
obeys the logarithmic law. 
In nontrivial NESSs far from equilibrium,  
$S_L$ is enhanced anomalously 
to obey the quasi volume law (\ref{eq:W>0,Dm>0}).
Consequently, 
$S_L > O(S_L^{\rm eq})$
in contrast to the finding in previous works \cite{eisler2005entanglement,aschbacher2007non,hoogeveen2015entanglement} that $S_L \leq O(S_L^{\rm eq})$
in NESSs. 
This anomalous behavior arises from the far from equilibrium property and 
multiple scatterings due to impurities
that break the translational symmetry of the system.
This suggests that 
similar results may be obtained for 
other models with certain symmetry-breaking scatterers,
 although we have studied only one such model in this paper.

\begin{acknowledgments}
We thank Sho Sugiura, Chisa Hotta, Yusuke Kato, 
T. Sagawa, E. Iyoda, N. Shiraishi,
Mamiko Tatsuta, and 
Hosho Katsura for valuable discussions.
This work was supported by The Japan Society
for the Promotion of Science, KAKENHI No. 26287085 and No. 15H05700.
\end{acknowledgments}

%


\clearpage

\pagestyle{empty}
\parindent=0mm
\setlength{\textwidth}{1.15\textwidth}

\begin{figure*}
\vspace{-10mm}\hspace{-28mm}
\includegraphics[page=1,clip]{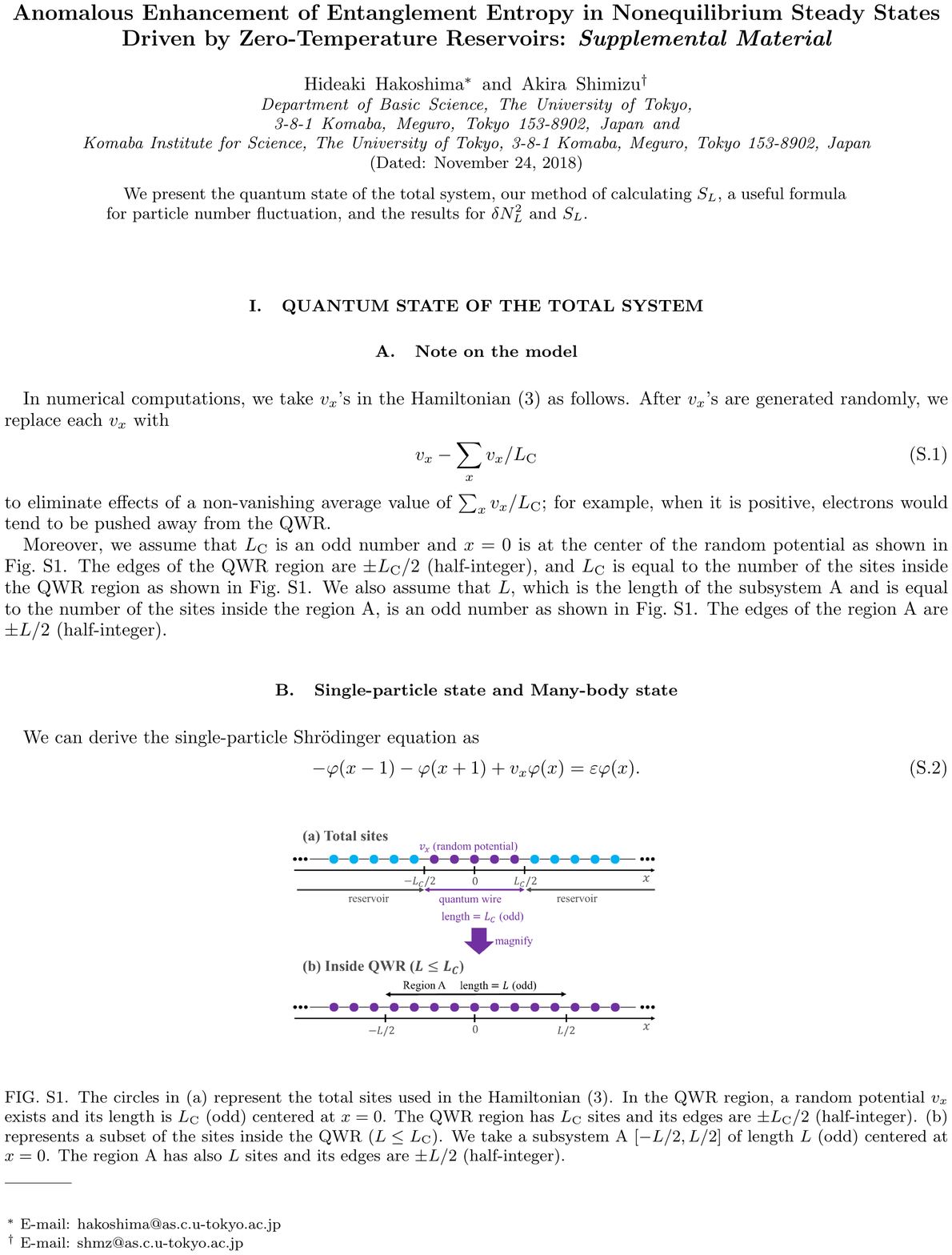}
\end{figure*}

\begin{figure*}
\vspace{-10mm}\hspace{-28mm}
\includegraphics[page=2,clip]{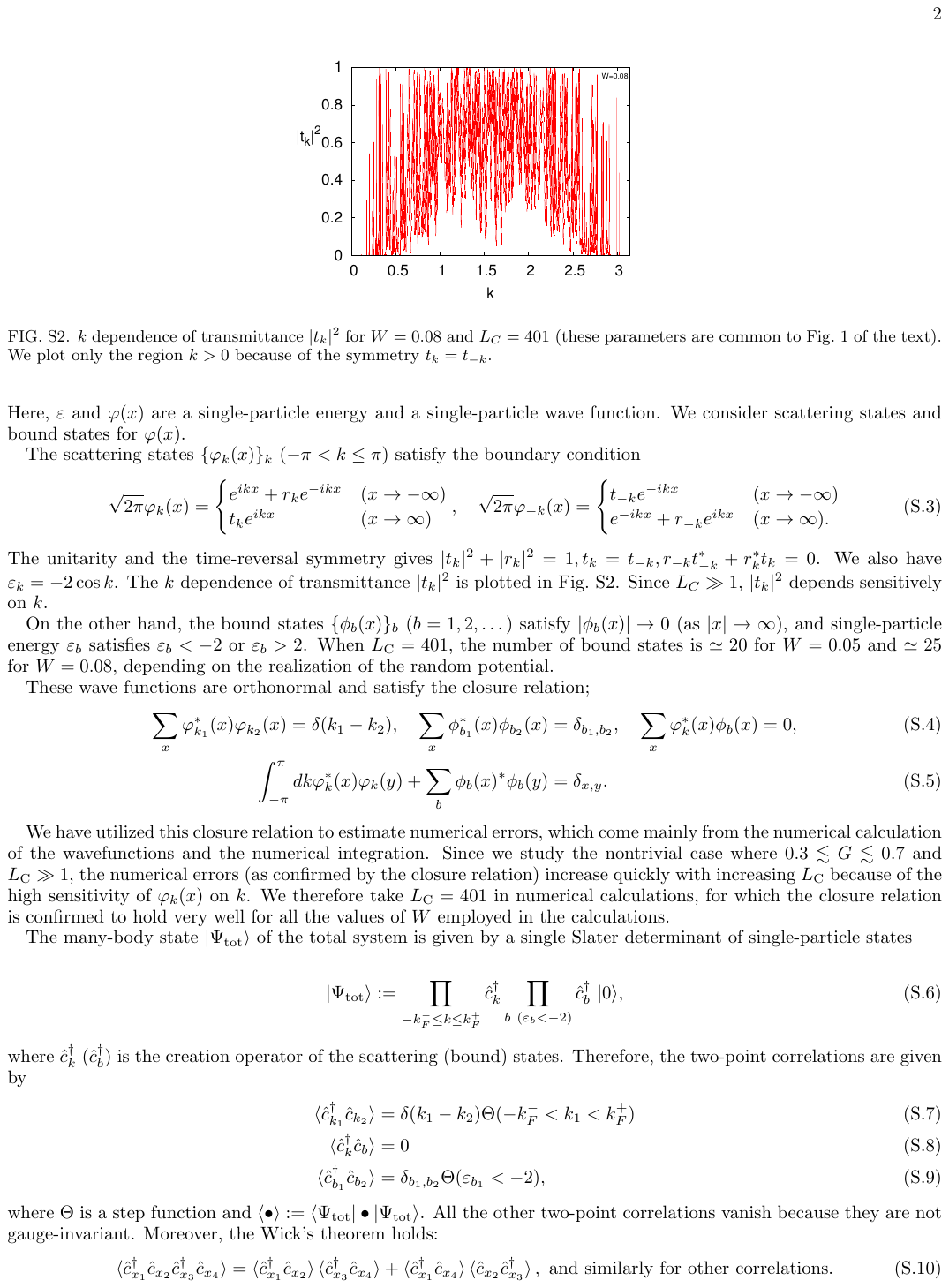}
\end{figure*}

\begin{figure*}
\vspace{-10mm}\hspace{-28mm}
\includegraphics[page=3,clip]{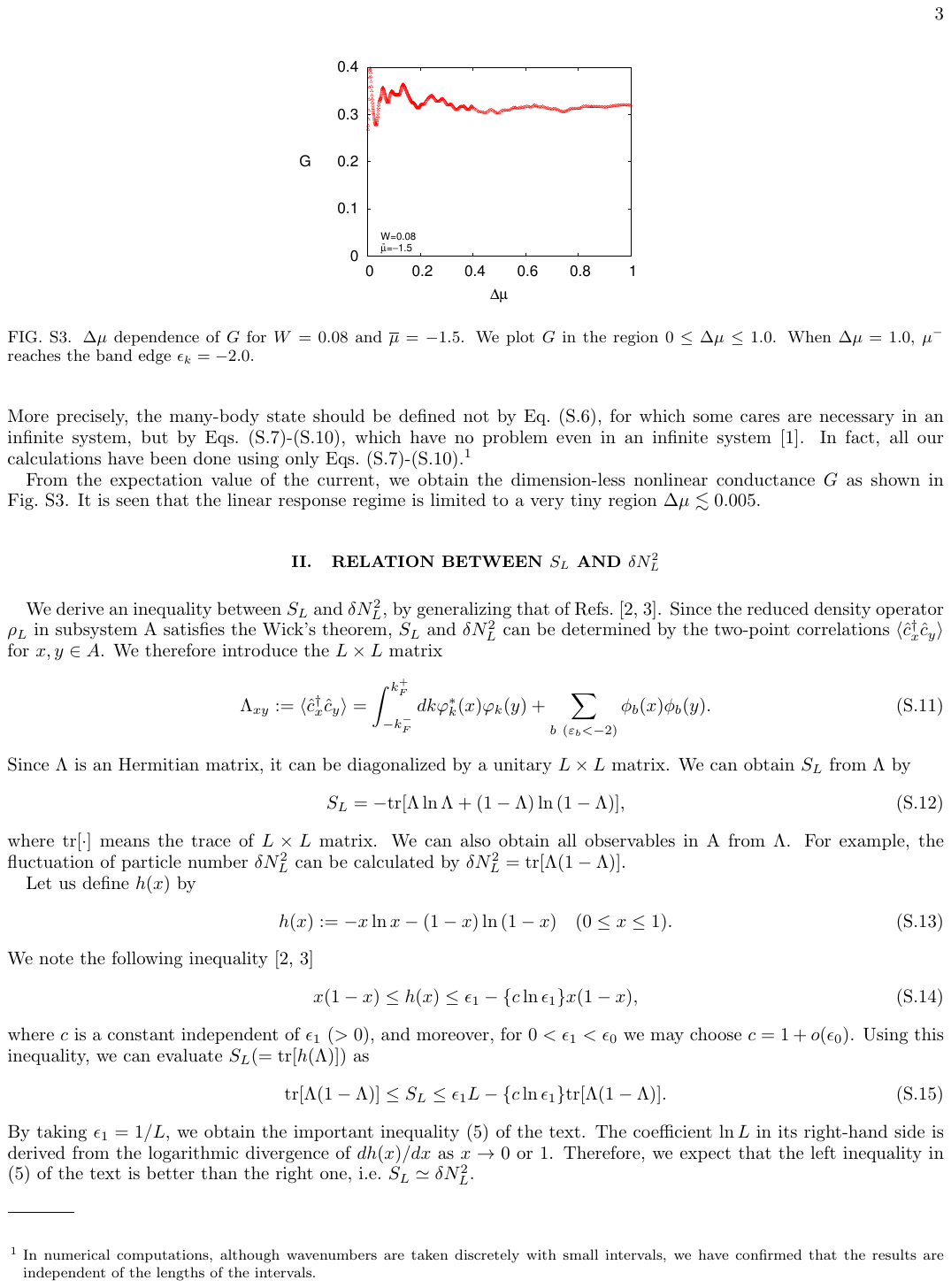}
\end{figure*}

\begin{figure*}
\vspace{-10mm}\hspace{-28mm}
\includegraphics[page=4,clip]{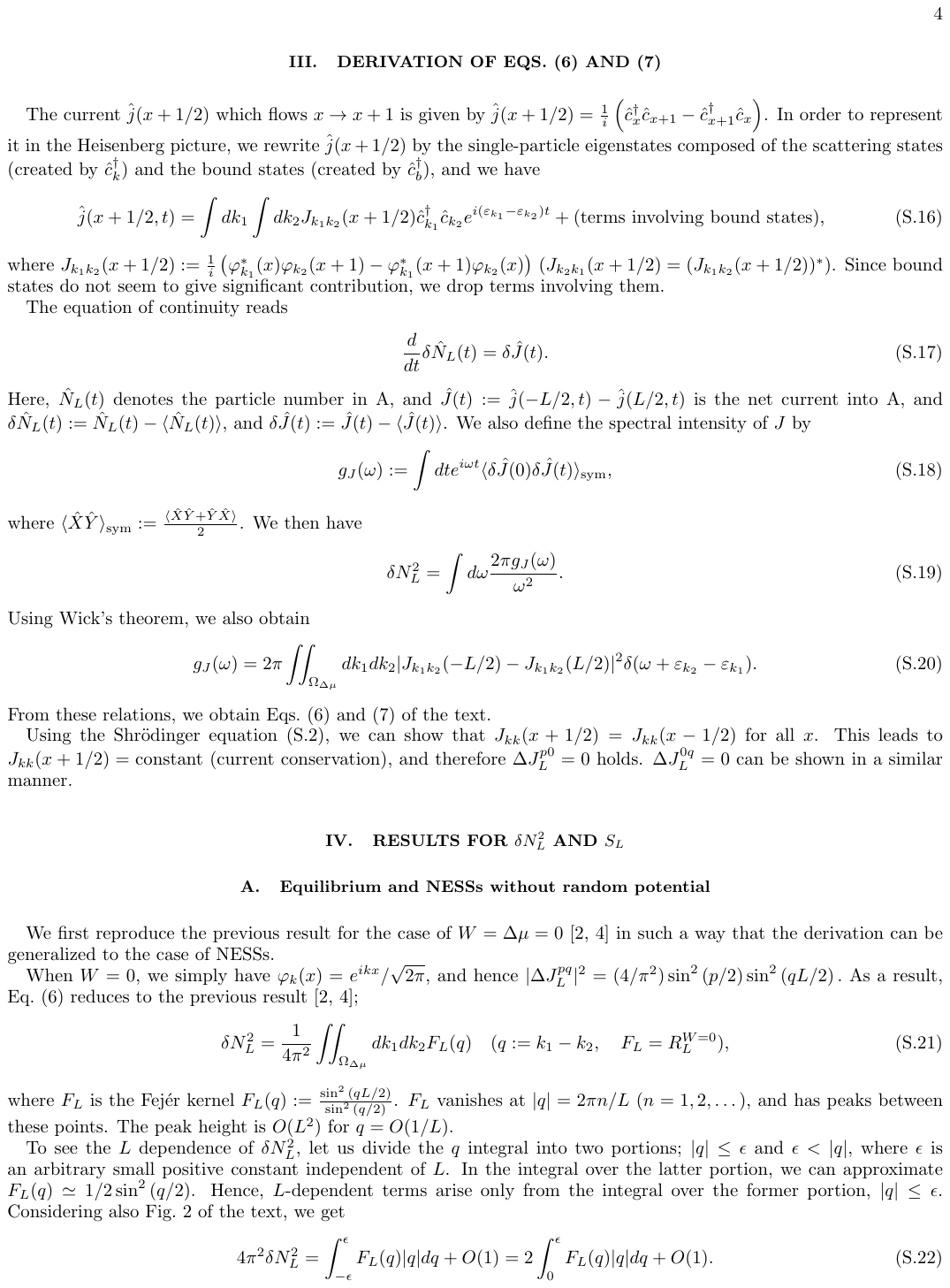}
\end{figure*}

\begin{figure*}
\vspace{-10mm}\hspace{-28mm}
\includegraphics[page=5,clip]{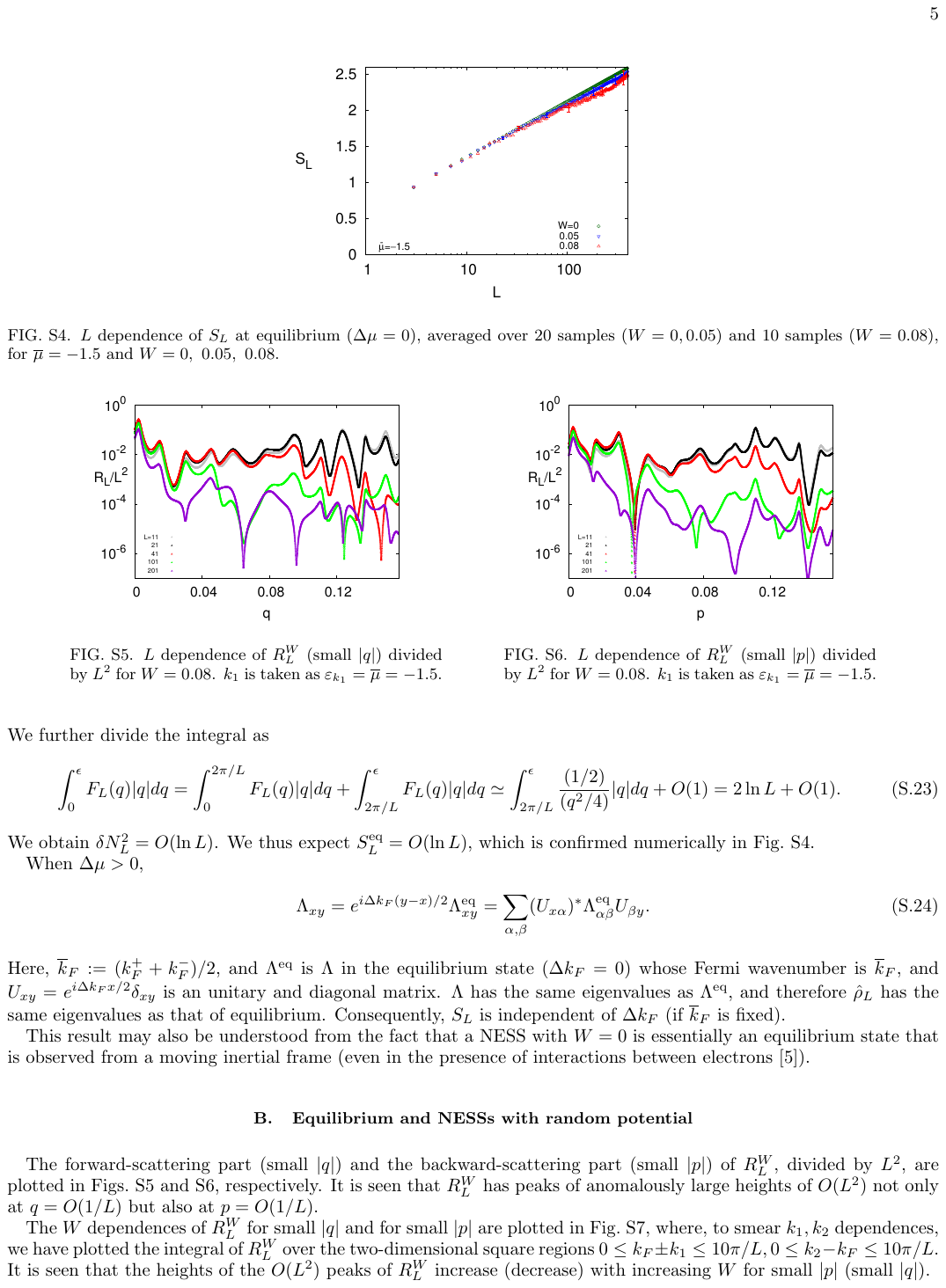}
\end{figure*}

\begin{figure*}
\vspace{-10mm}\hspace{-28mm}
\includegraphics[page=6,clip]{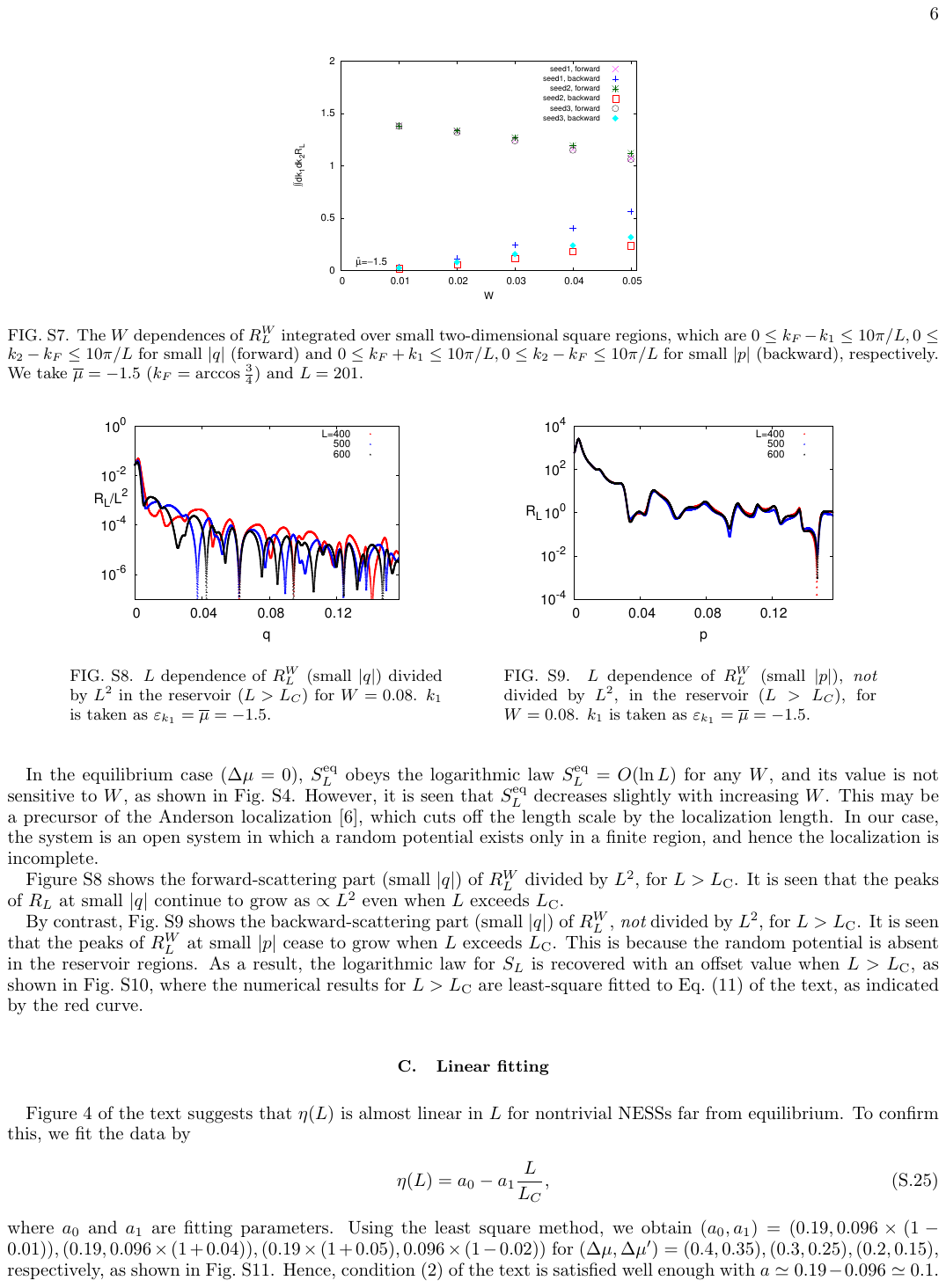}
\end{figure*}

\begin{figure*}
\vspace{-10mm}\hspace{-28mm}
\includegraphics[page=7,clip]{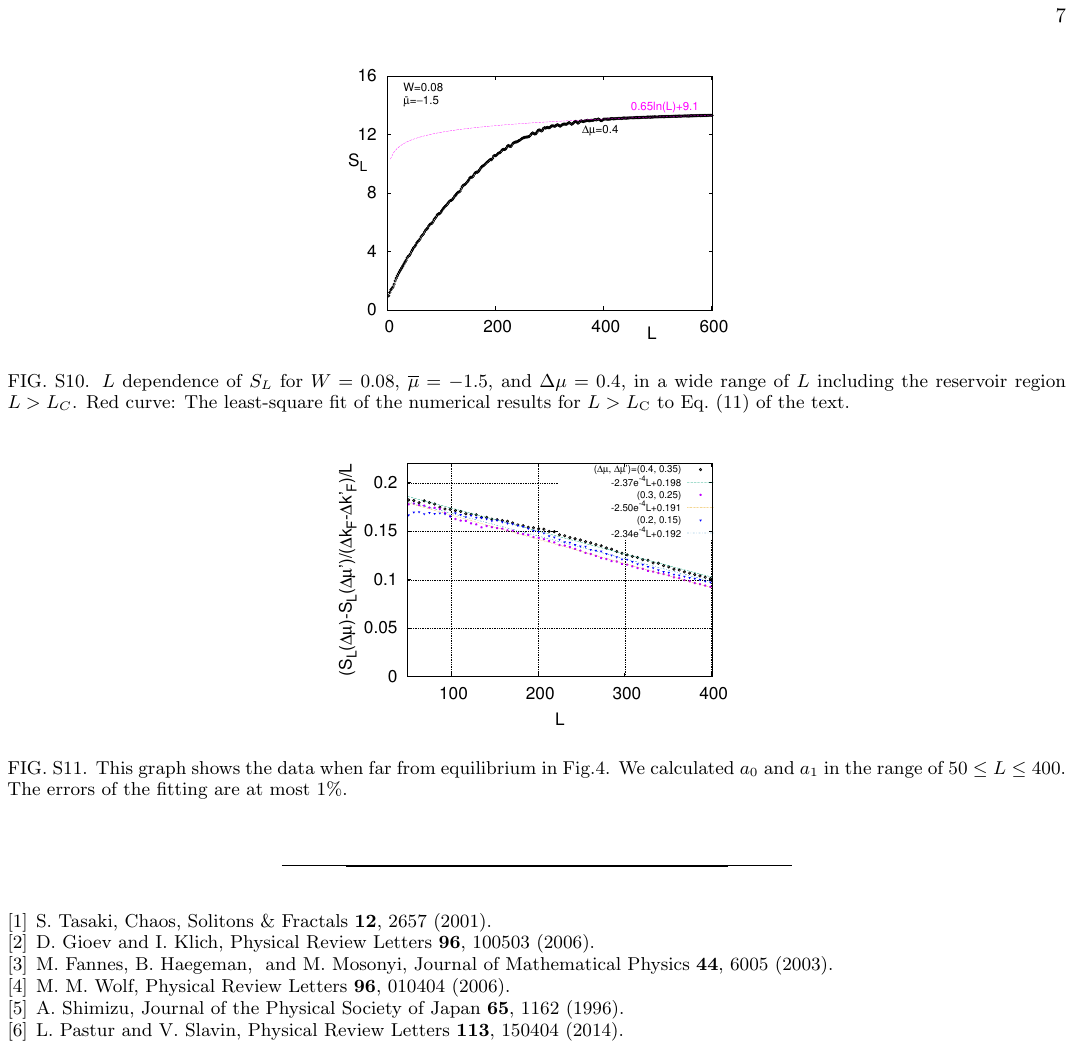}
\end{figure*}



\end{document}